# Tuning Coherent Radiative Thermal Conductance in Multilayer Photonic Crystals

Wah Tung Lau, Jung-Tsung Shen, Georgios Veronis, and Shanhui Fan[a]

*Ginzton Laboratory, Stanford University, Stanford, California, 94305, USA*

Paul V. Braun

*Department of Materials Science and Engineering, University of Illinois at Urbana-Champaign, Urbana, Illinois 61801, USA*

December 31, 2007

We consider coherent radiative thermal conductance of a multilayer photonic crystal. The crystal consists of alternating layers of lossless dielectric slabs and vacuum, where heat is conducted only through photons. We show that such a structure can have thermal conductance below vacuum over the entire high temperature range, due to the presence of partial band gap in most of the frequency range, as well as the suppression of evanescent tunneling between slabs at higher frequencies. The thermal conductance of this structure is highly tunable by varying the thickness of the vacuum layers.

[a] Electronic mail: shanhui@stanford.edu

Recently there has been significant interest in exploring coherent thermal transport, especially for devices at nanoscale.[1-8] To demonstrate coherent thermal transport, most experiments have exploited phonons as the heat carriers, and were typically performed at low temperature in order to avoid strong phonon-phonon interaction in solids.[1-6] On the other hand, interaction between photons is much weaker. It would therefore be of great interest as well to design structures that allow exploration of photon-based coherent thermal transport, which should occur at higher temperatures.

In this letter, we consider the thermal conductance of a multilayer photonic crystal, consisting of alternating layers of dielectric and vacuum (Fig. 1). In this structure, since neither electrons nor phonons can propagate through the vacuum layers, the thermal transport along the direction of periodicity should dominantly be due to photons. With the use of transparent materials as the dielectric, the absorption length of photons can be made to be much longer than the period of the crystal, and hence the heat conduction process should be coherent over significant distances. Moreover, unlike one-dimensional nanowire systems that have been extensively used to study coherent thermal transport,[7,8] here the photon modes are not confined to a narrow channel, and hence the photonic band structure in full three dimensions needs to be considered. Our analysis here shows that the thermal conduction properties of this system are significantly influenced by the band structure. In particular, the frequencies of the first bandgap in many cases define a characteristic temperature. At far below this characteristic temperature, the conductance of the crystal is the same as that of an equivalent uniform medium with an effective average index related to the dielectric constants of the layers. Above this characteristic temperature, throughout the entire temperature range, the conductance becomes smaller than that of vacuum, due to the presence of partial band gap in most of the frequency range, as well as the suppression of evanescent tunneling between slabs at higher frequencies. Since the band structure, which controls the characteristic temperature, is directly controlled by the geometry of the crystal, the conductance of the structure becomes highly tunable. As an example, we show that the conductance of this system can be changed by nearly an order of magnitude simply by altering the thickness of the vacuum layers.

As a starting point, we consider a photonic crystal described by a band structure $\omega_j(\mathbf{k})$ where $\omega$ is the photon frequency, $\mathbf{k}$ is the photon wavevector, and $j$ is the band index which includes both polarizations. Assuming that the heat flows along the z-direction, the thermal conductance per unit area is given by:[9]

$$G(T) = \sum_j \int_{k_z \geq 0} \frac{d^3\mathbf{k}}{(2\pi)^3} \hbar\omega_j(\mathbf{k}) \left|\frac{\partial \omega_j(\mathbf{k})}{\partial k_z}\right| \frac{\partial}{\partial T}\left(\frac{1}{e^{\hbar\omega_j(\mathbf{k})/(k_B T)} - 1}\right). \qquad (1)$$

This equation represents a generalization of the one-dimensional result in Ref. 1. We note that

$$\sum_j \int_{k_z \geq 0} \frac{d^3\mathbf{k}}{(2\pi)^3} \left|\frac{\partial \omega_j}{\partial k_z}\right| f(\omega_j) = \int_0^\infty \frac{d\omega}{2\pi} f(\omega) \left[\sum_m \int_{\text{constant } \omega} \frac{d^2\mathbf{k}_\parallel}{(2\pi)^2}\right], \qquad (2)$$

where $m$ labels all possible solutions of $k_z > 0$, including both polarizations, for a given pair of $(\omega, \mathbf{k}_\parallel)$. Here $\mathbf{k}_\parallel$ is the wavevector component on the x-y plane. (Notice that $m$ and the band index $j$ are generally not the same.) Eq. (1) can be further simplified to yield:

$$G(T) = k_B \int_0^\infty \frac{d\omega}{(2\pi)^3} \frac{[\hbar\omega/(k_B T)]^2 e^{\hbar\omega/(k_B T)}}{[e^{\hbar\omega/(k_B T)} - 1]^2} A(\omega), \qquad (3)$$

where

$$A(\omega) \equiv \sum_m \int_{\text{constant } \omega} d^2\mathbf{k}_\parallel. \qquad (4)$$

At a constant frequency $\omega_0$, $\omega_0 = \omega_j(\mathbf{k})$ describes constant frequency surfaces in $\mathbf{k}$-space. $A(\omega = \omega_0)$ is the total projected area of such constant frequency surfaces onto the x-y plane, and will be referred to as the "projected area" hereafter in this paper. $A(\omega)$ arises because photons are allowed to propagate in any direction. Thus, in three-dimensional structures, the dispersion relations of photons can play a significant role in determining thermal conductance. This is in contrast to one-dimensional structures such as nanowires where the carriers propagate strictly in one dimension. In one-dimensional systems, if one further assumes that there is only one mode that contributes to thermal conductance, at low enough temperature the thermal conductance assumes a universal form $(\pi k_B^2 T)/(6\hbar)$ that is independent of the dispersion relation.[1]

In order to highlight the effects of band structure on thermal conductance, below for most cases we will normalize the conductance of the photonic crystals against that of vacuum. For vacuum in three dimensions, with $\omega(\mathbf{k}) = c|\mathbf{k}|$, one has $A_{vac}(\omega) = 2\pi(\omega/c)^2$ and $G_{vac}(T) = (\pi^2 k_B^4 T^3)/(15\hbar^3 c^2)$. Note that $A_{vac}(\omega)$ and $G_{vac}(T)$ include contributions from both polarizations.

Using Eqs. (3) and (4), we now consider the properties of thermal conductance of the multilayer structure shown in Fig. 1. We assume that the dielectric slabs in Fig. 1 are lossless, with a thickness of $d_1$ and a refractive index $n_1 = n$. The vacuum layer has a thickness $d_2$ and a refractive index $n_2 = 1$. $a = d_1 + d_2$ is the periodicity. For such a multilayer geometry, for each value of $(\omega, \mathbf{k}_\parallel)$, there are two independent polarization modes: the *s*-polarized mode, with electric field along the direction of $\hat{\mathbf{z}} \times \mathbf{k}_\parallel / k_\parallel$; and the *p*-polarized mode, with magnetic field along the direction of $\hat{\mathbf{z}} \times \mathbf{k}_\parallel / k_\parallel$. Here $\mathbf{k}_\parallel$ is conserved in all layers, and $k_\parallel = |\mathbf{k}_\parallel|$. The dispersion relations for both polarizations are given by:[10]

$$\cos(k_z a) = \cos(\gamma_1 d_1)\cos(\gamma_2 d_2) - \frac{1}{2}\left(\frac{\gamma_1}{P_\zeta \gamma_2} + \frac{P_\zeta \gamma_2}{\gamma_1}\right)\sin(\gamma_1 d_1)\sin(\gamma_2 d_2). \quad (5)$$

Here $\gamma_{i=1,2} = \sqrt{(n_i \omega/c)^2 - k_\parallel^2}$. $P_s = 1$ and $P_p = n^2$ for the *s*- and *p*-polarized fields respectively. At any given $\omega$, a real solution for $k_z$ represents a Bloch state of propagating mode of the photons. These solutions of $(\omega, k_\parallel, k_z)$ define the photonic bands. The collection of all Bloch states in the $\omega - k_\parallel$ plane defines the projected band diagram.[10] As an example, the projected band diagram for *s*-polarization is shown in Fig.

2(a). The refractive index of the slab is taken as $n = \sqrt{12}$, which approximates the refractive index of silicon at infrared frequencies and at room temperature.

In the projected band diagram, the Bloch states exist only in the region $k_\| \leq n\omega/c$, i.e. only in the region below the light line of the dielectric. Among all Bloch states of the system, the states in the region below the vacuum light line, (i.e. $k_\| \leq \omega/c$) are extended in both the dielectrics and vacuum. The states in the region between the vacuum and the dielectric light lines, (i.e. $\omega/c \leq k_\| \leq n\omega/c$) are extended in the dielectrics but evanescent in vacuum.

With the projected band diagram as defined by Eq. (5), one can compute the projected area $A(\omega)$ using Eq. (4). Fig. 2(b) shows the projected area $A_s(\omega)$ for $s$-polarized modes, normalized with respect to $A_{vac}(\omega) = 2\pi(\omega/c)^2$. Here $A_s(\omega) = 2\pi \sum_m \int_{k_{m,s}^{(1)}(\omega)}^{k_{m,s}^{(2)}(\omega)} k_\| dk_\|$, where $k_{m,s}^{(1)}(\omega)$ and $k_{m,s}^{(2)}(\omega)$ are the minimum and maximum values of $k_\|$ of the $m^{th}$ $s$-polarized band at $\omega$. Since $A_{vac}(\omega)$ contains contributions from both polarizations, for vacuum $A_s(\omega)/A_{vac}(\omega) = 1/2$. For the crystal, at $\omega \to 0$, $A_s(\omega)/A_{vac}(\omega) = \frac{1}{2}\frac{n_1^2 d_1 + n_2^2 d_2}{d_1 + d_2} > \frac{1}{2}$ as shown in Fig. 2(b). The photonic crystal thus behaves as a uniform medium with an effective dielectric constant. In contrast, over the entire high frequency range [Fig. 2(b)], $A_s(\omega)/A_{vac}(\omega) < 1/2$. As the frequency increases, the bandwidth of the modes above the vacuum light line ($k_\| > \omega/c$) becomes progressively smaller. These modes are evanescent in vacuum, and as the frequency increases, the evanescent decay lengths in vacuum approach zero. [Fig. 2(a)]. Consequently their contributions to $A_s(\omega)$ gradually vanish. Below the vacuum light line ($k_\| < \omega/c$), partial bandgaps persist. Thus, $A_s(\omega)/A_{vac}(\omega) < 1/2$ in the high frequency region. The transition of between the low and high frequency behavior typically occurs within the first bandgap, provided that at a frequency in this gap, the evanescent lengths in vacuum of the modes become comparable to the thickness of the vacuum layer. The *p*-polarization, and hence $A(\omega) \equiv A_s(\omega) + A_p(\omega)$, exhibit the same transition behaviors from low to high frequency region.

Having obtained $A(\omega)$, we now calculate $G(T)$ using Eq. (3). The result, normalized with respect to $G_{vac}(T)$, is shown in Fig. 3, along with the respective contributions from *s*- and *p*- polarizations. The curve of $G(T)/G_{vac}(T)$ as a function of $T$ generally follows the shape of the curve $A(\omega)/A_{vac}(\omega)$ in $\omega$, with fluctuations smoothened by averaging over the energy distribution of photons. The total normalized thermal conductance of the structure $G(T)/G_{vac}(T)$ (black solid curve) is much larger than 1 at low temperatures, and decreases to a value smaller than 1 at higher temperatures. This transition can be understood from the behavior of $A(\omega)$. At low temperature, most photons populate a small bandwidth of frequency from $\omega = 0$, where $A(\omega)/A_{vac}(\omega) > 1$, thus $G(T)/G_{vac}(T) > 1$. At high temperature, photons spread over to a wide range of frequency, where $A(\omega)/A_{vac}(\omega) < 1$. As a result, $G(T)/G_{vac}(T) < 1$. Thus, photonic crystals can be used to generate a medium with total thermal conductance significantly below that of the vacuum.

Since the band structure is directly determined by geometry, the use of photonic crystal provides a mechanism to drastically tune the thermal conductance at a constant temperature. As an example, we vary the vacuum spacing between the dielectric slabs for the structure shown in Fig. 1. Increasing the spacing decreases the frequency of the first bandgap. Consequently, one could tune the structure from $G < G_{vac}$ to $G > G_{vac}$ at selected temperatures. As an example, in Fig. 4, we plot the total thermal conductance as a function of $T$ for $d_2 = 9d_1$, $d_1$ and $0.111d_1$. In this parameter range of $d_2$, with fixed thickness $d_1$ of the dielectric slab, the total thermal conductance decreases as the spacing $d_2$ increases. In particular, when the operating temperature is at $T_a = 0.02hc/(k_B d_1)$, the increase of $d_2$ results in the decrease of total thermal conductance by more than an order of magnitude (as shown by the inset in Fig. 4).

To observe the coherent photon thermal transport effects predicted in this paper, we note that the frequency range of integration in Eq. (3) is set by the function $k_B \frac{[\hbar\omega/(k_B T)]^2 e^{\hbar\omega/(k_B T)}}{[e^{\hbar\omega/(k_B T)}-1]^2}$, which approximately vanishes at $\hbar\omega > 10 k_B T$. Thus at a given temperature $T$, the material chosen needs to be transparent in the energy range $[0, 10 k_B T]$. For temperature at $300K$, this range is $[0, 0.259 eV]$. We note that intrinsic silicon in fact is transparent in this entire range,[11] because optical phonon in silicon is not optically active. Using silicon as the dielectric layer to observe the tunable thermal conduction effect as plotted in Fig. 4, we can choose $d_1 = 1\mu m$. The choice of the temperature $T_a$ that has large tunability then corresponds to $288K$. Thus, such an effect can be potentially observed at room temperature.

**Reference**

[1] K. Schwab, E. A. Henriksen, J. M. Worlock, and M. L. Roukes, Nature (London) **404**, 974 (2000).

[2] L. G. C. Rego and G. Kirczenow, Phys. Rev. Lett. **81**, 232 (1998).

[3] T. Yamamoto, S. Watanabe, and K. Watanabe, Phys. Rev. Lett. **92**, 075502 (2004).

[4] G. Chen, J. of Heat Transfer **121**, 945 (1999).

[5] M. V. Simkin and G. D. Mahan, Phys. Rev. Lett. **84**, 927 (2000).

[6] A. N. Cleland, D. R. Schmidt, and C. S. Yung, Phys. Rev. B **64**, 172301 (2001).

[7] M. Meschke, W. Guichard, and J. P. Pekola, Nature (London) **444**, 187 (2006).

[8] D. R. Schmidt, R. J. Schoelkopf, and A. N. Cleland, Phys. Rev. Lett. **93**, 045901 (2004).

[9] J. Ziman, *Electrons and Phonons* (Oxford University Press, Oxford, England, 1960).

[10] A. Yariv and P. Yeh, *Photonics: Optical Electronics in Modern Communications* (Oxford University Press 6$^{th}$ edition, 2007).

[11] E. D. Palik, *Handbook of Optical Constants of Solids* (Academic Press, 1985), p. 554.

**Figure captions**

FIG. 1. Schematic of the geometry. Gray layers are dielectric slabs. The slabs are separated by vacuum.

FIG. 2. (a) The projected band diagram for the $s$-polarized modes for the structure shown in Fig. 1, with $d_1 = 0.5a$. The bands in the region $0 < k_\| < \omega/c$ are drawn in brown; and those in the region $\omega/c < k_\| < n\omega/c$ are drawn in green. The vacuum $k_\| = \omega/c$ and the material $k_\| = n\omega/c$ light-lines are also shown. The inset shows the projection of the constant frequency surface on the $x$-$y$ plane for $\omega = 0.61(2\pi c/a)$.

(b) The blue line shows $A_s(\omega)/A_{vac}(\omega)$ for $s$-polarization. The brown and green lines represent the contributions in the range $0 < k_\| < \omega/c$, and $\omega/c < k_\| < n\omega/c$ respectively.

(c) The black, blue, and red lines are the total $A(\omega)$, $A_s(\omega)$, and $A_p(\omega)$, respectively, as normalized to $A_{vac}(\omega)$.

FIG. 3. Normalized radiative thermal conductance $G(T)/G_{vac}(T)$ for the structure shown in Fig. 1 with $d_1 = 0.5a$. The contributions of $s$-polarized (blue curve) and $p$-polarized (red curve) modes are summed to give the total conductance (black curve). The inset shows a wider range of temperature.

FIG. 4. Normalized radiative thermal conductance $G(T)/G_{vac}(T)$ as one varies slab separation $d_2$. The temperature of $T_a = 0.02 hc/(k_B d_1)$, where significant variation as a function of $d_2$ occurs, is indicated as a vertical dotted line. The inset shows $G(T)/G_{vac}(T)$ at $T_a = 0.02 hc/(k_B d_1)$, as $d_2$ varies continuously from $d_2 = 0.111 d_1$ to $d_2 = 9 d_1$.

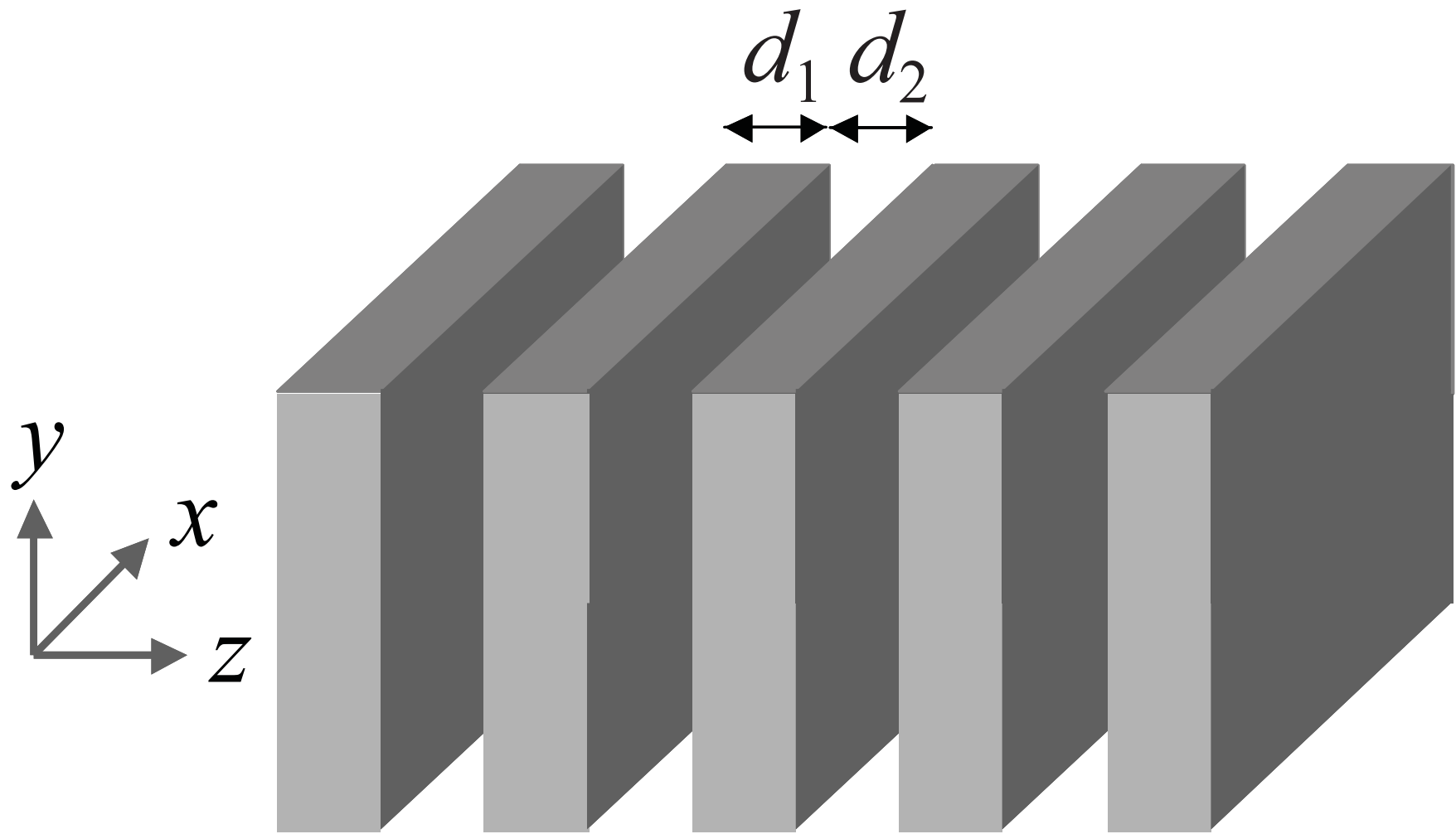

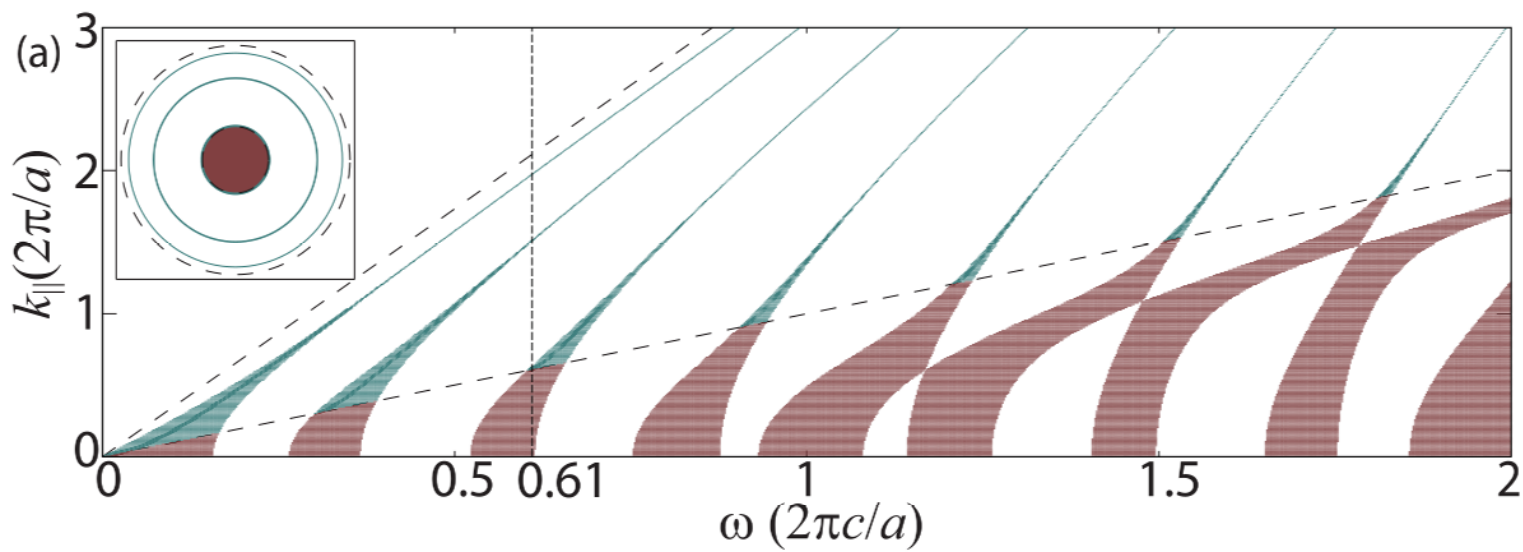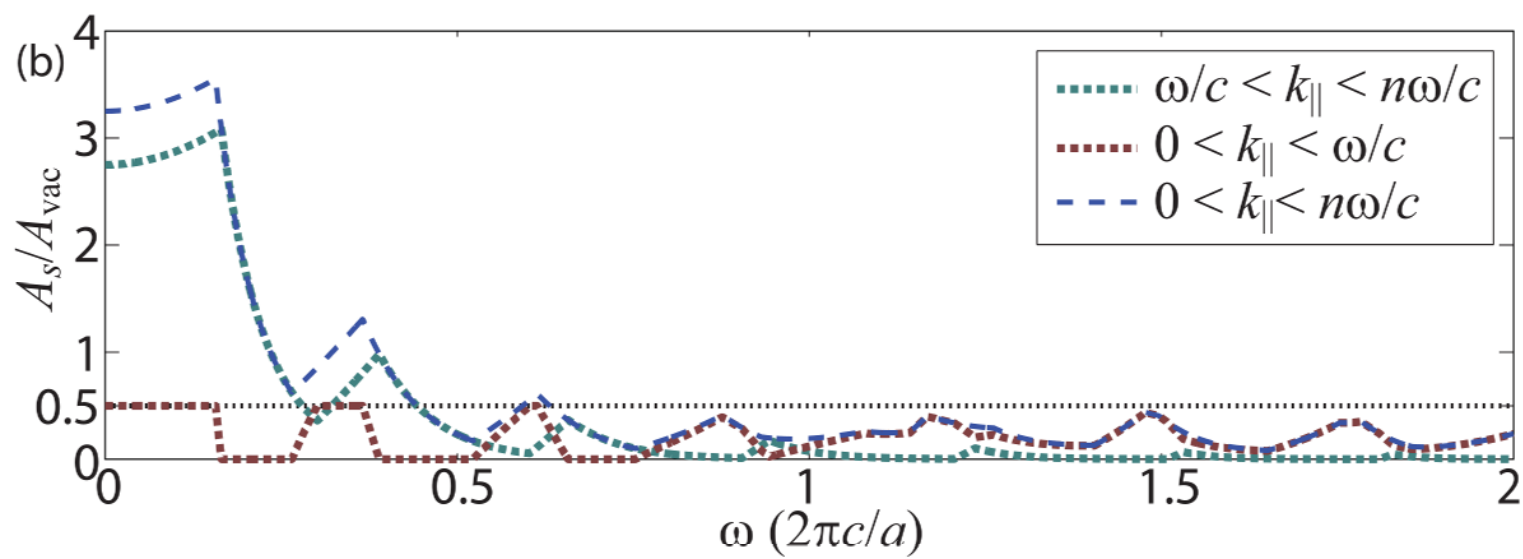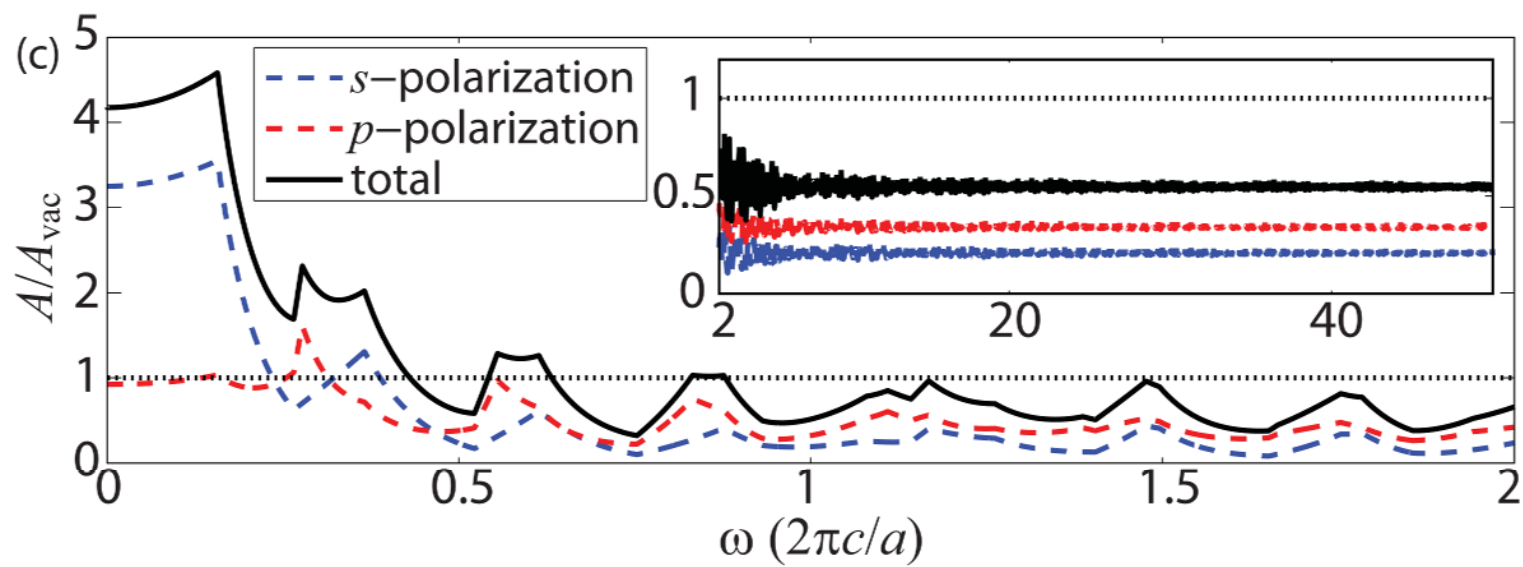

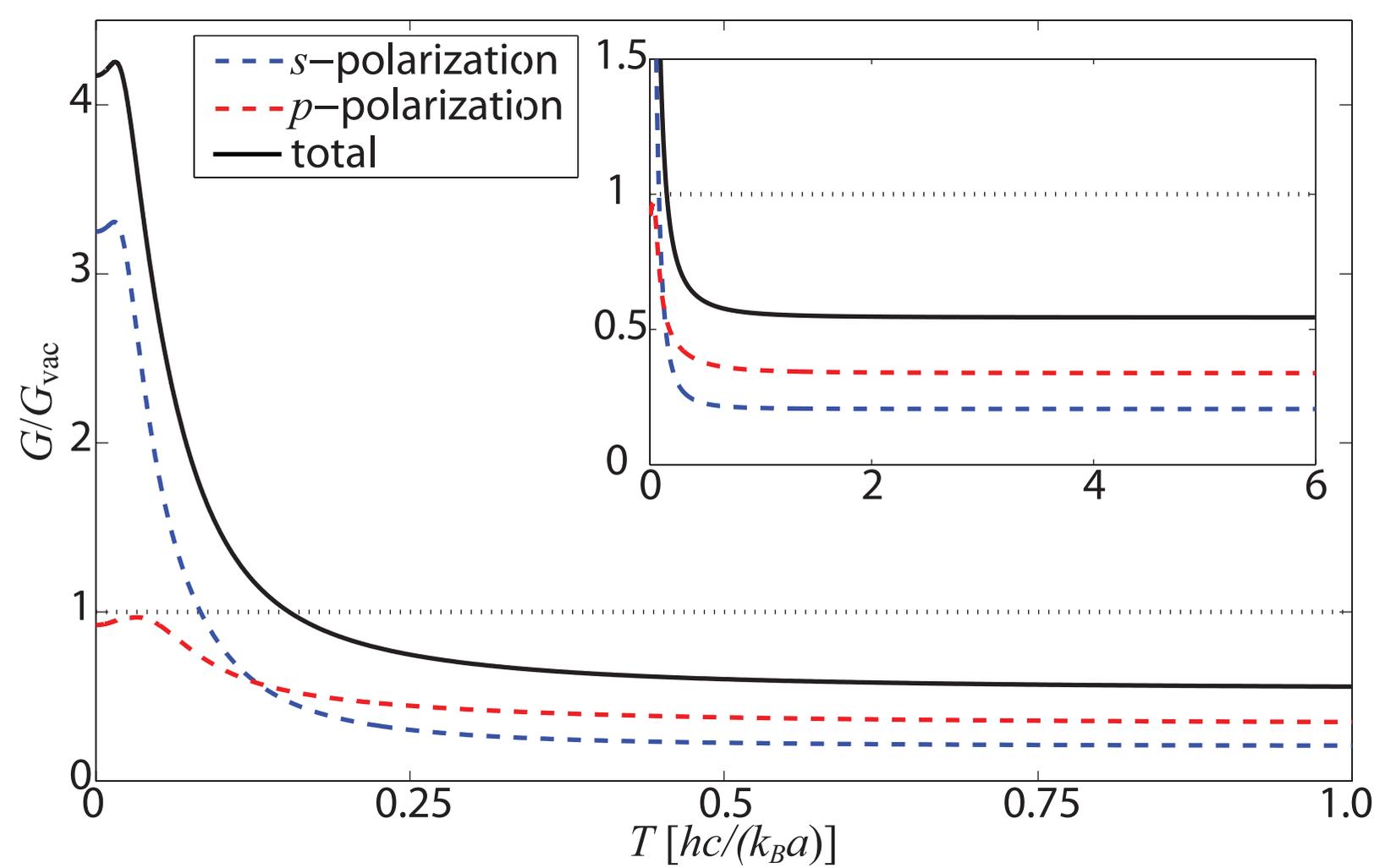

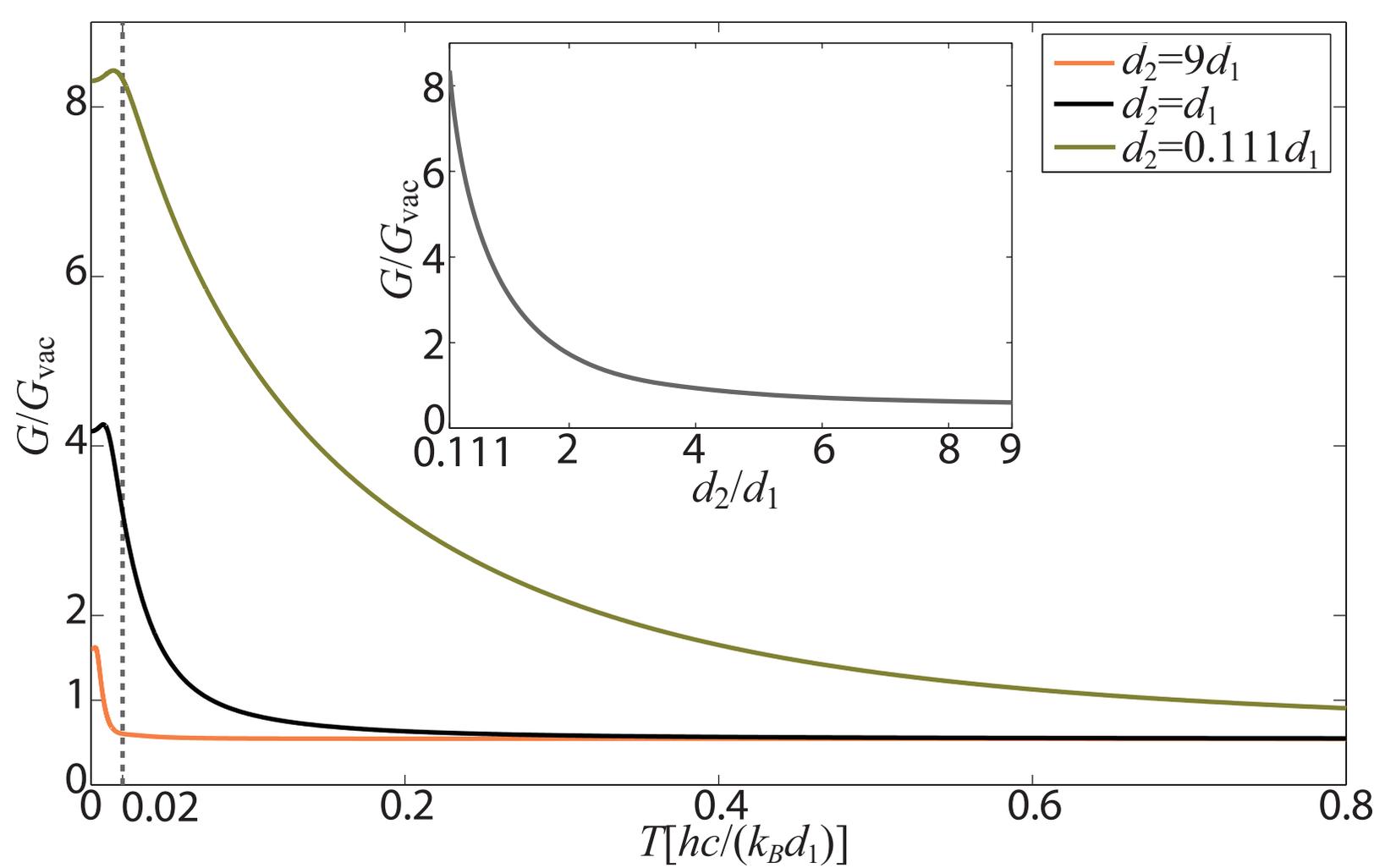